\documentclass[9pt]{article}
\newcommand{\figref}[2]{Figure \ref{#1}{\bf{}#2}}%
\usepackage{graphicx}
\usepackage[letterpaper,left=1in,right=1in,top=1in,bottom=1in]{geometry}
\usepackage{authblk}
\date{}
\begin{document}

\title{Of Matters Condensed}
\author{Michael Shulman\footnote{shulman@physics.harvard.edu} }
\author{Marc Warner}
\affil{Department of Physics, Harvard University, \\Cambridge, MA 02138}
\maketitle

\begin{abstract}

The American Physical Society (APS) March Meeting of condensed matter
physics has grown to nearly 10,000 participants, comprises 23
individual APS groups, and even warrants its own hashtag
($\#$apsmarch). Here we analyze the text and data from March Meeting
abstracts of the past nine years and discuss trends in condensed
matter physics over this time period. We find that in comparison to
atomic, molecular, and optical physics, condensed matter changes
rapidly, and that condensed matter appears to be moving increasingly
toward subject matter that is traditionally in materials science and
engineering.

\end{abstract}

\section*{Introduction} 
Condensed matter physics studies beautiful fundamental physics, such
as superconductivity, while simultaneously providing important
inventions including the transistor, CCD, integrated circuit and diode
laser. Moreover, the field holds promise for future applications, such
as dissipationless power transfer and quantum information
processing. The American Physical Society (APS) March Meeting is
arguably the largest condensed matter physics conference and attracts
nearly 10,000 participants to a single location to discuss the
topic. These participants include undergraduate students, graduate
students, postdocs, staff scientists, professors, engineers, vendors
of scientific instruments, and journalists both from around the
world. Sessions include numerous contributed talks, posters, invited
talks, (so-called ``crackpot'' sessions), and exhibits, and provides
all physics enthusiasts an opportunity to deepen their knowledge of
their respective fields of specialization, as well as to gain exposure
to new, emerging fields in condensed matter. The March Meeting,
therefore, offers a snapshot of progress in the represented fields,
and is a good measure of the trends in condensed matter. We scraped
the abstract content from the March Meeting programs from 2007-2015 to
examine and analyze these trends, and speculate about the future of
matters condensed.

\section*{Some simple numbers} 

\begin{figure}
{\includegraphics[scale=0.5]{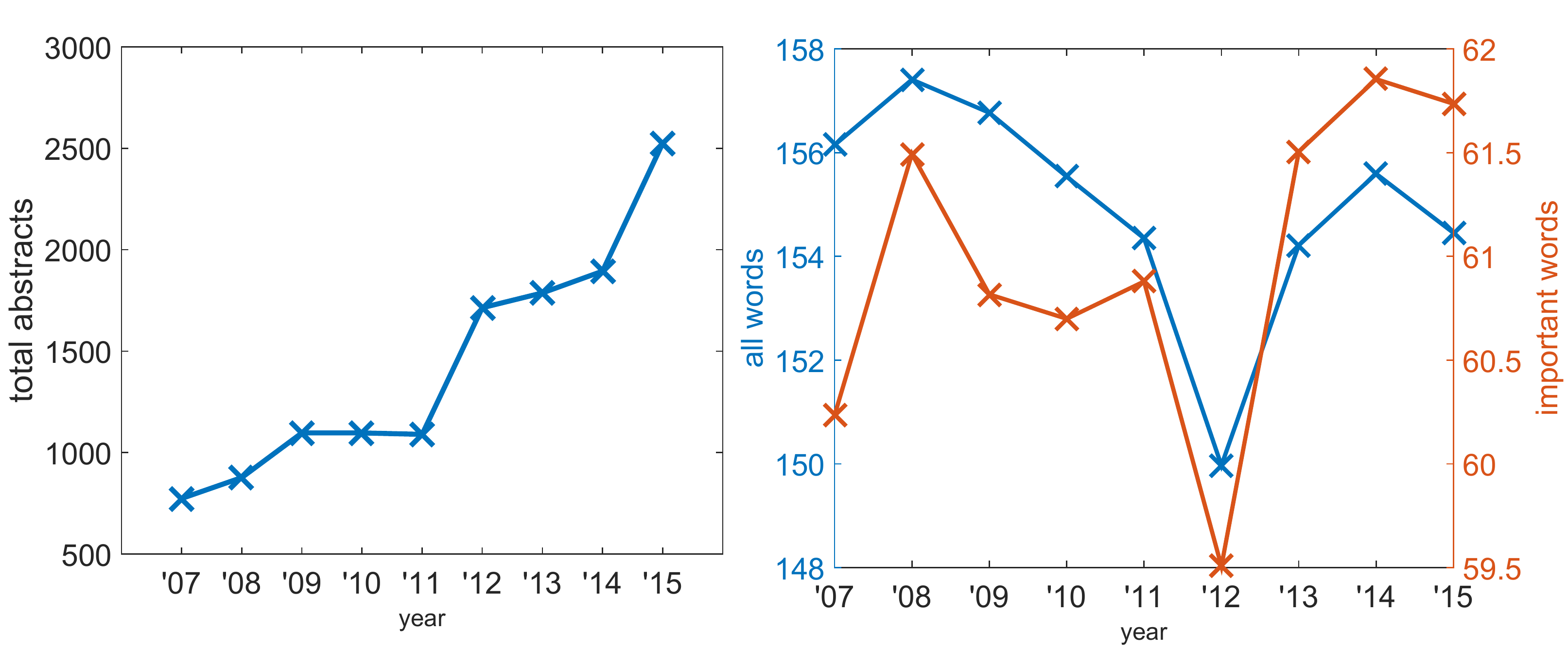}}
\caption{\textbf{a:} The number of abstracts submitted to March
  Meeting has steadily risen since 2007. \textbf{b:} The total words
  per abstract and important words per abstract have remained
  approximately constant. }
\label{numAb}
\end{figure}

We begin with the most basic of trends. The field of condensed matter,
as measured by the number of contributed abstracts to the March
Meeting has shown steady growth since 2007 (\figref{numAb}{a}). This
finding is somewhat surprising, given the sentiment that interest in
condensed matter was inflated by the discoveries of the 50's, 60's and
70's, and has subsequently begun to wane. In fact, as measured by
arxiv submissions, the growth of condensed matter has slowed, and its
relative contribution to the entire arxiv pool has steadily dropped
since approximately
2002\footnote{http://arxiv.org/help/stats/2014\_by\_area/index}. The
steady growth of abstracts in the presence of a decrease in arxiv
submissions can be attributed either to submissions of historically
``cond-mat'' articles to other areas (such as ``quant-phys''), or to
an increase in unpublishable abstract submissions. However, our
expertise is in only a small subset of March Meeting and condensed
matter fields, and we therefore cannot adequately judge the
publishability of results. Lastly, we note remarkable consistency in
both the length (in words) and number of important keywords (as
defined below) in each abstract (\figref{numAb}{b}), suggesting that
contributors have continued to obey the APS abstract guidelines, and
tend to the same point in the tradeoff between readability and
information density.

\section*{Single word analysis} We employ standard procedures used in natural
language processing to analyze the text. We treat each abstract in the
``bag of words'' approximation, using only word frequency, and
neglecting word order and context. Additionally, we ``clean'' text,
stripping away suffixes to more accurately group together similar
words, creating a dictionary of 24000 words. Lastly, we ignore common
words, and normalize all word counts by the total number of abstracts
in a given year. This creates a document matrix for each March Meeting
from 2007 to 2015, containing the percentage of abstracts which
contain a specific word.

\begin{figure} 
\begin{center}
{\includegraphics[scale=0.5]{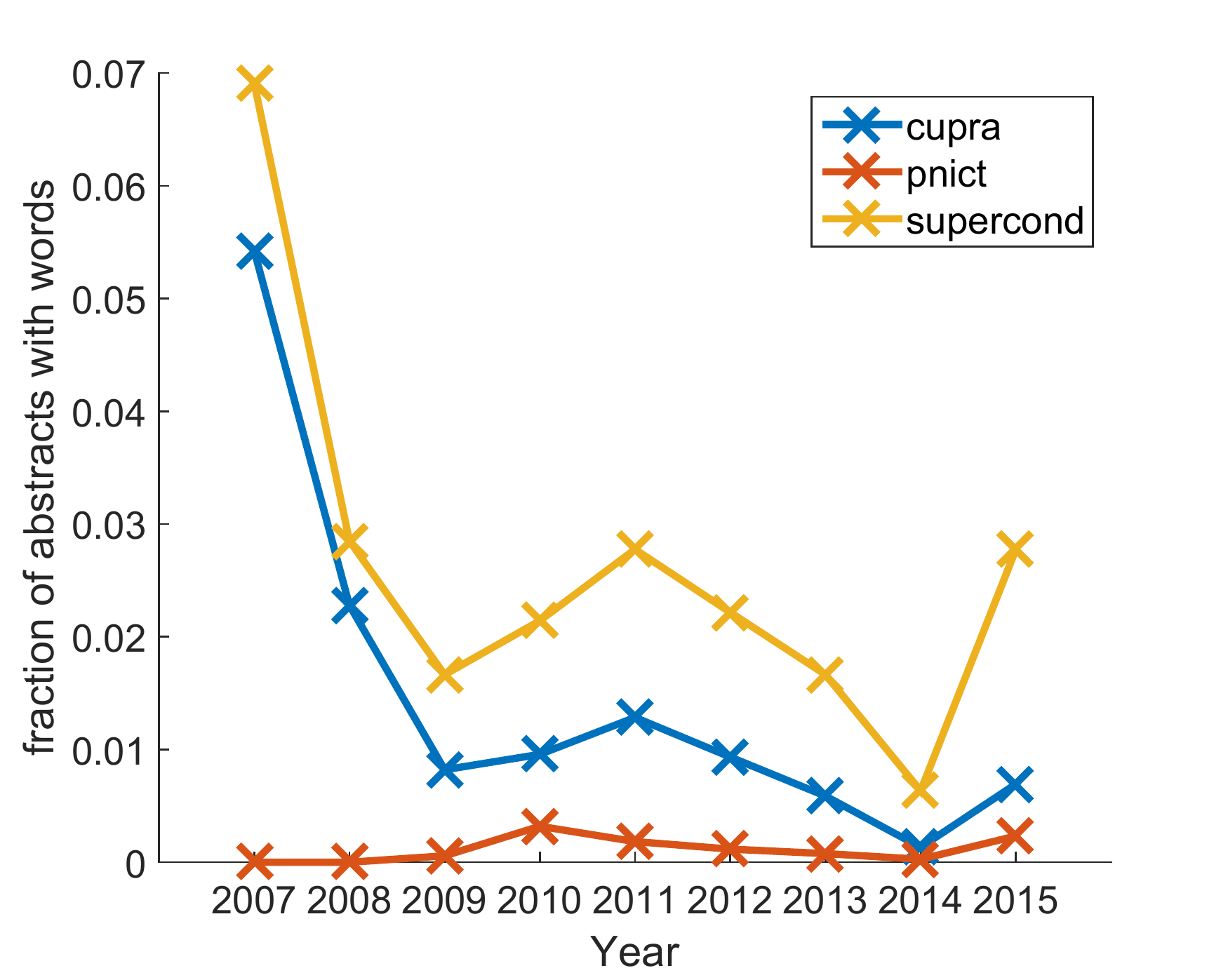}} 
\end{center}
\caption{The trends in popularity of various kinds of high-temperature
  superconductors are largely dominated by the decline in popularity of
  superconductivity itself. }
\label{supercond} 
\end{figure}

We use these word counts to analyze various trends in March Meeting
abstracts. For example, we can see the rise and fall of various
families of high-temperature superconductor, though these trends are
indeed dominated by an ``envelope'' of dimishing interest in
superconductivity (\figref{supercond}{}).  We attribute the rise in
abstracts which reference superconductivity to the recent interest in
topological superconductivity and Majorana Fermions.

\begin{figure} 
\begin{center}
{\includegraphics[scale=0.5]{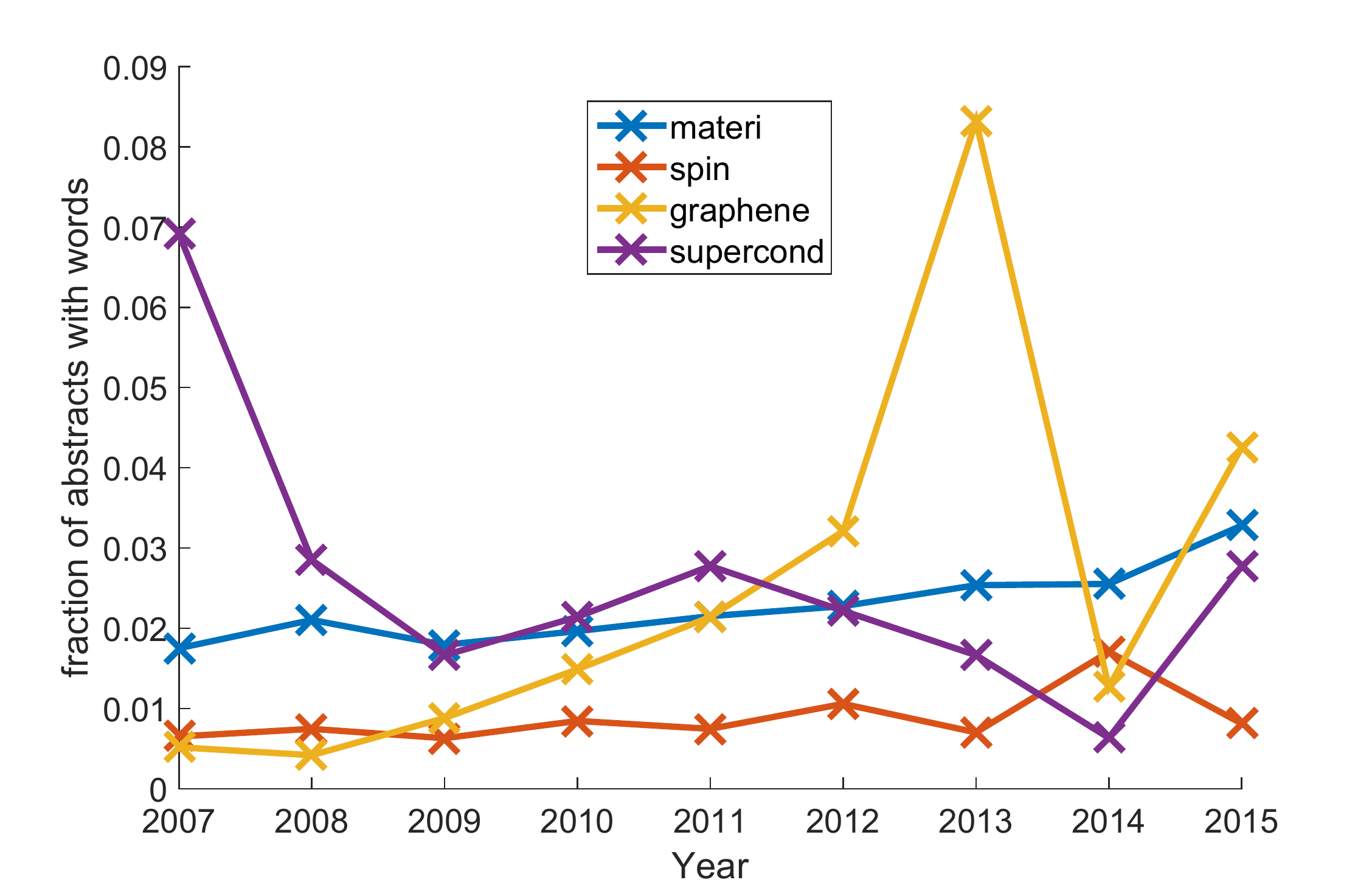}} 
\end{center}
\caption{While superconductivity shows nearly steady decline, other
  topics such as materials show steady growth in popularity.}
\label{material} 
\end{figure}

We gain further insight into the direction of condensed matter physics by
eximining this decline in popularity of superconductivity, with the voltile
behavior of graphene, and the steady trend toward studying interesting
materials (\figref{material}{});

\section*{Big movers} 

\begin{figure}
\begin{center}
{\includegraphics[scale=0.5]{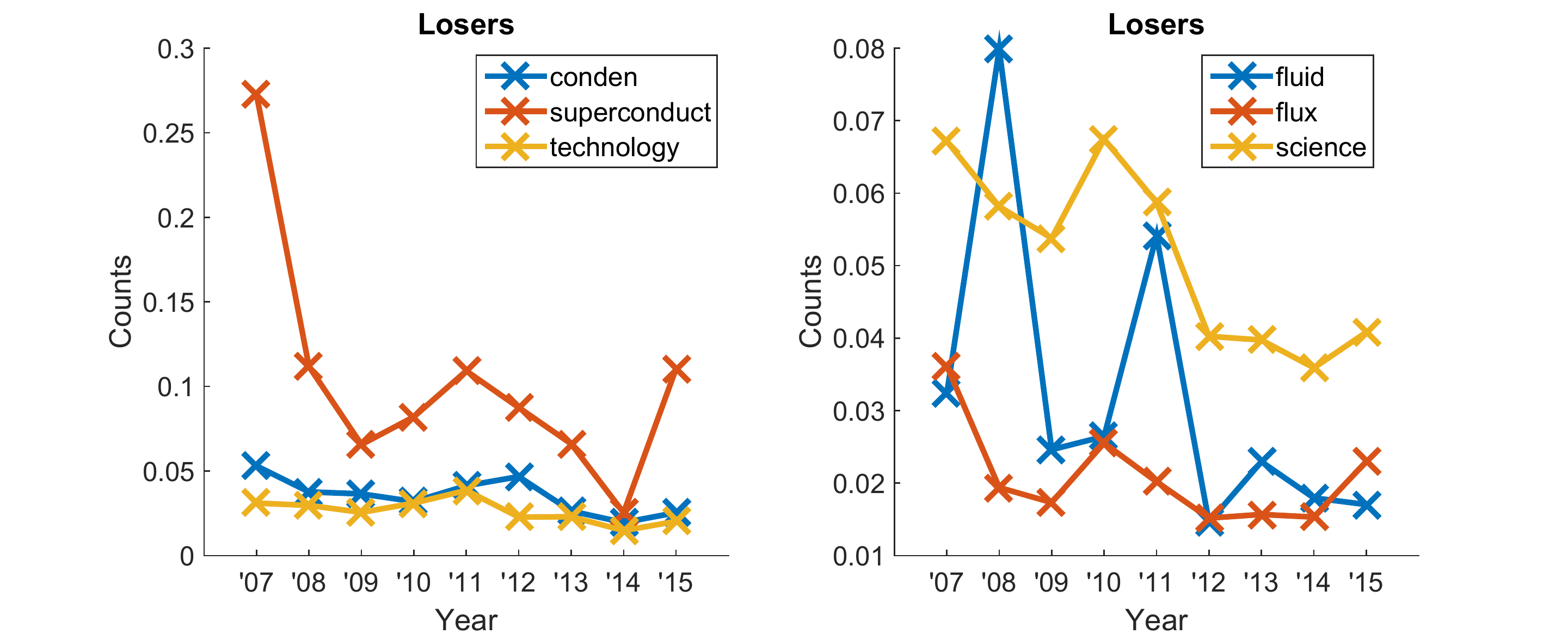}}
\end{center}
\caption{Some surprising (e.g. ``science'') and unsurprising (e.g. ``superconduct'') salient words with large losses in popularity over the last nine years.}
\label{bigLose}
\end{figure}

We choose the salient words to examine by finding the words which are
among the 1000 most popular keywords in all the years. We then compute
the slope of line ($\frac{dP_{word}}{dt}$) for each of the feature
words, where $P_{word}$ is the fraction of abstract with a particular
word, and we identify the ``big movers'' as the keywords with very
large (positive or negative) slopes. In \figref{bigLose}{} we plot the
trends of some of the relevant big losers of the plast nine years,
including superconductivity, which has likely seen a descreasing
interest due to difficulties with high-temperature
superconductivity. There are some more surprising terms with
declining popularity over the past nine years, including ``conden''
and ``science,'' though it is unclear why. We leave the reader to draw
additional conclusions and to make further speculations based on these
data.

\begin{figure}
\begin{center}
{\includegraphics[scale=0.5]{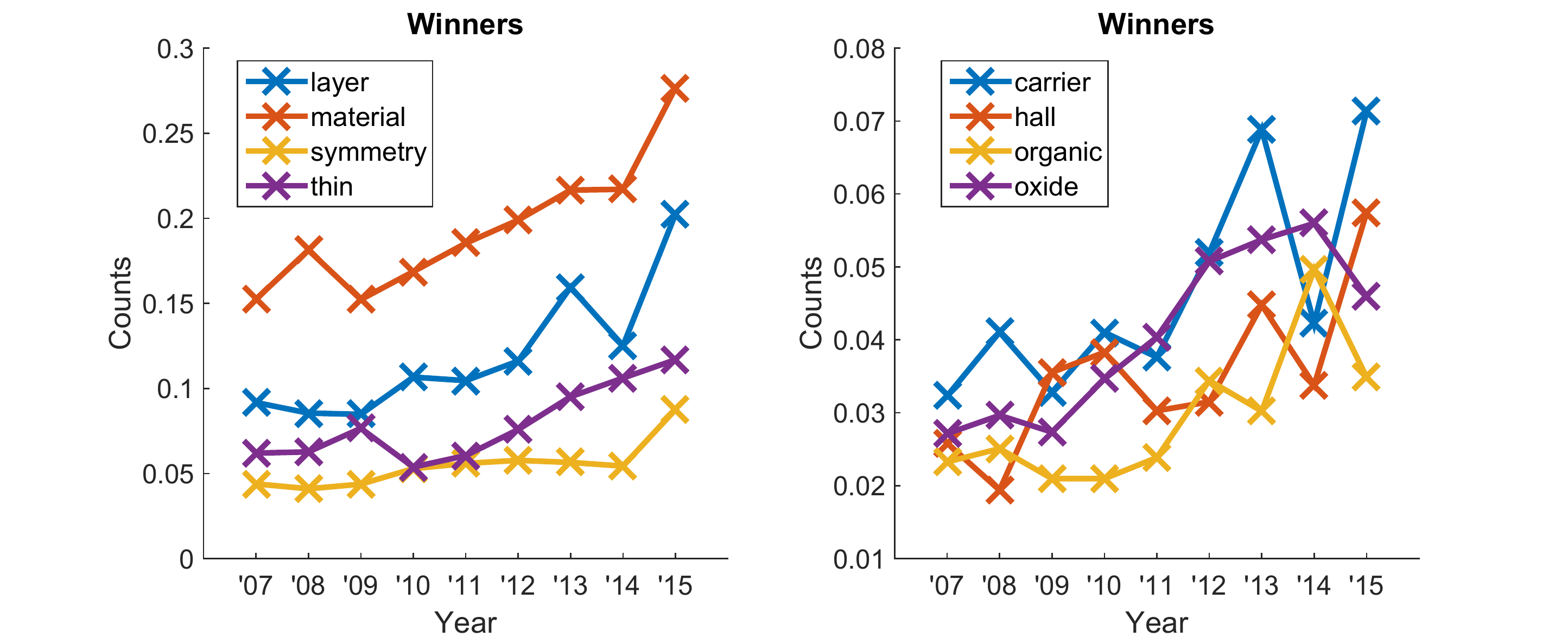}}
\end{center}
\caption{Some salient words with large gains in popularity. These are
  largely associated with material science, suggesting that the field
  of condensed matter is moving toward subject matter that is
  typically associated with material science.}
\label{bigWin}
\end{figure}

We plot some of the terms with the largest increases in popularity over the
past several years in \figref{bigWin}{}. The subject matter with large
increases largely pertain to interesting materials and material
properties, and these trends appear to continue, or perhaps strengthen, up to
2015.

\section*{Latent semantic analysis}
We confirm these observed trends, as well as make more quantitative
claims using latent semantic analysis. We create a sparse
$N_{abstract}\times{}N_{words}$ matrix for each year containing word
counts for each abstracts. Using singular value decomposition we
reduce the dimentionality of these matricies and find the salient
word-feature vectors. In 2007, for example, the words with the most
weight in the feature vector with the largest singular value in
descending order are:
\begin{verbatim}
{supercond, dop, gap, magne, wave, curpa, state, pair, low, meas,
 high, spin, delta, show, stat, CuO, tunnel, electro, phase}
\end{verbatim}
which we recognize as representing ``high-temperature
superconductivity.'' Similarly, for 2014, we find the highest weight
words in the feature with the largest singular value to be:
\begin{verbatim}
{magnet, spin, field, ferro, temperature, order, meas, coupl, aniso,
 antiferro, lattice, low, exchang, orbit, frustr, neutron, momentum}
\end{verbatim}
which we identify as ``novel magnetic materials'' and for 2015 we find
\begin{verbatim}
{film, propert, sub, layer, thin, high, micro, materi, depo, appli,
 grow, spectrum, device, nano, meas, ray, surface, studi}
\end{verbatim}
which we idenity as ``novel layered materials.'' Indeed, our
suspicions of trends in subject matter away from high-temperature
superconductivity toward topics which border matieral science are
confirmed.

For comparison, we can repeat the same analysis on abstracts from the
division of atomic, molecular and optical physics (DAMOP), and we find
much more uniformity among feature vectors from year to year. For
example, the most significant words of the feature vector with the
largest singular value for 2007 are
\begin{verbatim}
{optic, quantum, atom, phase, trap, field, condens, lattice, two,
 demon, magne, cool, frequenc, pair, ultra, light}
\end{verbatim}
and for 2014\footnote{the 2015 abstracts are not yet available} are
\begin{verbatim}
{optic, quantum, spin, lattice, ultra, trap, atom, condens, magne,
 progra, phase, many, body, dimen}
\end{verbatim}
There is significant overlap between these two vectors (the years
between 2007 and 2014 are also similar), and both could be classified
as ``quantum optics with ultra-cold atoms.''




\section*{Outlook and conclusion} Despite what the arxiv submissions might
suggest, it appears that condensed matter is vibrant and growing, at
least in the short term. However, it appears that it adapts to the
pervading climate of the times, and the subject matter has evolved
from ``traditional'' condensed matter physics such as
superconductivity, to subject matter on the border with material
science. As \figref{bigLose}{} and the latent semantic analysis
suggests, condensed matter may be less focused on the ``condensed,''
but has branched out into the more applied field of materials
science. We see no signs of this trend stopping (nor any reason to
resist it) and we speculate that as funding and interest correlate
with applications, that the field will continue to trend to the more
applied.

\section*{Acknowledgements}
The authors acknowledge John Nichol for fruitful discussions. 
\end{document}